# Analog pile-up circuit technique using a single capacitor for the readout of Skipper-CCD detectors

**M. Sofo Haro,**[a,b,c,d,1] **C. Chavez,**[a,f] **J. Lipovetzky,**[b,c,d] **F. Alcalde Bessia,**[b,c,d] **G. Cancelo,**[a] **F. Chierchie,**[b,g] **J. Estrada,**[a] **G. Fernandez Moroni,**[a] **L. Stefanazzi,**[a] **J. Tiffenberg,**[a] **S. Uemura,**[e]

[a] *Fermi National Accelerator Laboratory,*
  *Batavia, Illinois, USA.*

[b] *Consejo Nacional de Investigaciones Científicas y Tecnicas (CONICET),*
  *Argentina.*

[c] *Comisión Nacional de Energía Atómica (CNEA),*
  *San Carlos de Bariloche, Río Negro, Argentina.*

[d] *Centro Atómico Bariloche, Instituto Balseiro,*
  *San Carlos de Bariloche, Río Negro, Argentina.*

[e] *School of Physics and Astronomy, Tel-Aviv University,*
  *Tel-Aviv, Israel.*

[f] *Facultad de Ingeniería, Universidad Nacional de Asunción,*
  *Asunción, Paraguay.*

[g] *Instituto de Investigaciones en Ingeniería Eléctrica "Alfredo C. Desages", Departamento de Ingeniería Eléctrica y de Computadoras,*
  *Bahia Blanca, Argentina.*

  E-mail: miguelsofoharo@gmail.com

ABSTRACT: With Skipper-CCD detectors it is possible to take multiple samples of the charge packet collected on each pixel. After averaging the samples, the noise can be extremely reduced allowing the exact counting of electrons per pixel. In this work we present an analog circuit that, with a minimum number of components, applies a double slope integration (DSI), and at the same time, it averages the multiple samples producing at its output the pixel value with sub-electron noise. For this prupose, we introduce the technique of using the DSI integrator capacitor to add the skipper samples. An experimental verification using discrete components is presented, together with an analysis of its noise sources and limitations. After averaging 400 samples it was possible reach a readout noise of $0.2\ e^-_{RMS}/pix$, comparable to other available readout systems. Due to its simplicity and significant reduction of the sampling requirements, this circuit technique is of particular interest in particle experiments and cameras with a high density of Skipper-CCDs.

---

[1]Corresponding author.

# Contents

# 1 Introduction

Charge Coupled Devices (CCD) has been the preferred detector of optical photons in ground and space based astronomy and in scientific laboratory applications. Due to their low readout noise, during the last decade they have been applied in dark matter (DM) and neutrino direct detection experiments [1] [2]. Both kind of experiments, use the CCD silicon mass as target material for the DM or neutrino particles.

Recently, the development of the Skipper-CCD detector has been a major technological breakthrough, producing an unprecedented reduction in the readout noise of CCDs, reaching $0.068\,e^-_{\text{rms}}/\text{pix}$ [3]. The sensor allows to reach the ultimate sensitivity of silicon material as a charge signal sensor by unambiguous determination of the charge signal collected by each cell or pixel, even for single electron-hole pair ionization.

Figure 1 is a simplified diagram of the output stage of a Skipper-CCD. Unlike conventional CCDs, Skipper-CCDs have a floating gate (FG) in the sensing stage connected to the gate of a source follower (SF) transistor [4][3]. Using a couple of additional gates (SG and OG in figure 1), the charge packet can be transferred below the FG multiple times, allowing sensing the charge non-destructively. No charge loss is produced in this process thanks to the buried channel [5]. When there is no charge below the FG, at the SF output there is the pedestal or baseline voltage level, and when the charge is transferred below FG there is the signal level. The difference between these two levels is proportional, by the sensitivity $S_v$ ($\mu V/e^-$), to the charge packet $Q$. If the charge samples are affected by uncorrelated noise, after averaging them, the noise falls with the square root of the number of samples [3].

The extremely low readout noise of Skipper-CCDs allowed the SENSEI experiment ("Sub-Electron Noise Skipper-CCD Experimental Instrument") to set world-leading constraints on DM-electron interactions for DM masses in the range of 500 keV to 5 MeV using a single Skipper-CCD [6–8]. Due to the discovery potential of this technology there are different experiments on progress. The SENSEI collaboration is planning to run an experiment with around 50 Skipper-CCD at SNOLAB. Also, the DAMIC-M (DArk Matter in CCDs at Modane) experiment will run an experiment with hundreds of Skipper-CCDs at the Modane underground lab [9]. Moreover, there is the OSCURA project that is developing the necessary technology for a dark matter experiment with thousands of Skipper-CCDs. The potential of Skipper-CCDs has also motivated its application in nuclear reactor neutrinos detection [10] [11]. These experiments requires a scalable and affordable readout system.

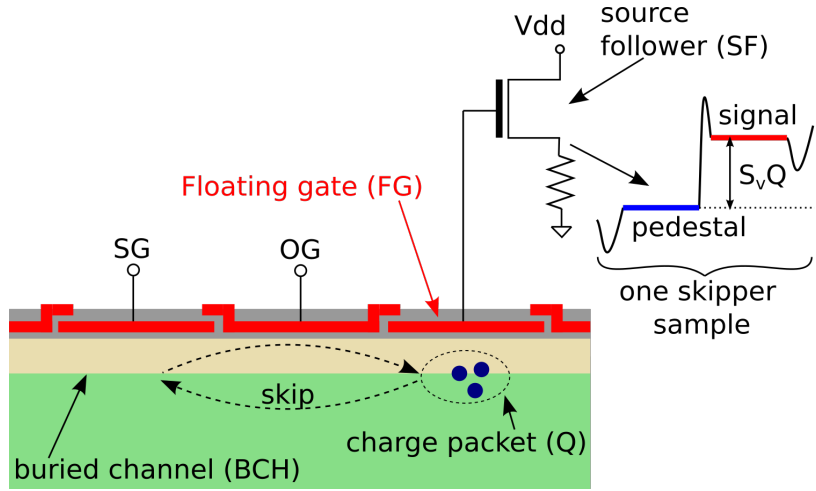

**Figure 1**. Simplified diagram of the output stage of a Skipper-CCD detector.

Recently, a specific readout system for Skipper-CCDs called LTA ("Low Threshold Acquisition") has been developed[12]. It is a digital-based system that integrates in a single board the processing of the Skipper-CCD video signal and the generation of the CCD control signals. The video signal is digitized by an ADC and processed in an FPGA [12]. Due to the digital nature of the system, it is hard to scale and costly. Also, radiopurity and low temperature operation limits its application in large scale DM experiment. In the following sections, we describe the development of an analog circuit, that we called Skipper-CDS, for the readout of Skipper-CCDs, that introduces a new technique that simplifies the architecture of a readout system with thousands of detectors.

## 2 Description and operation

As was mention in the introduction, a sample of the pixel charge is extracted from the difference between the pedestal and signal voltage levels, this technique is know as CDS ("Correlated Double Sampling") [13]. A widely used CDS is the DSI ("Dual Slope Integrator") because it provides an optimal filtering for white noise and completely rejects the reset noise [14] [15]. A deep theoretical analysis of the DSI filter can be found in [4] [16] [17] [12]. Equation 2.1 describes in time the operation of the DSI; $v_s(t)$ and $v_p(t)$ are the video signal during the pedestal and signal level respectively. Both are integrated during a time window $T$ and subtracted to obtain one sample of the pixel charge packet. All the skipper samples, $N_{skp}$, must be added to produce the pixel average value with sub-electron noise.

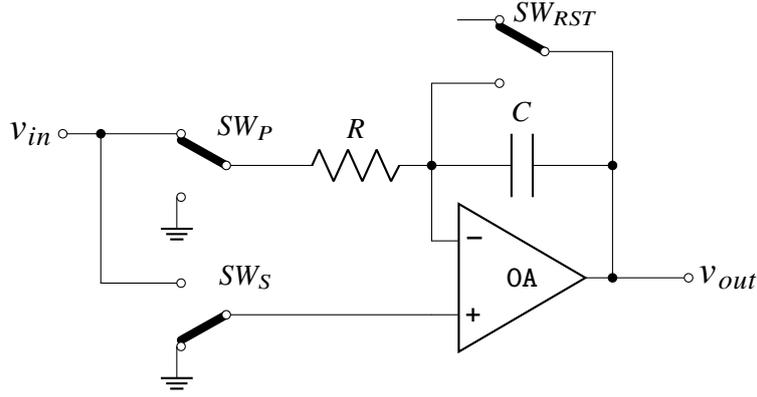

**Figure 2**. Skipper-CDS: Analog readout circuit proposed in this work for reading out Skipper-CCD detectors.

$$V_{pix} = \sum_{N_{skp}} \left[ \int_{<T>} v_s(t)dt - \int_{<T>} v_p(t)dt \right] \qquad (2.1)$$

Figure 2 is the circuit proposed in this work to perform the DSI operation for Skipper-CCDs. It is composed by an operational amplifier that, together with the resistor $R$ and capacitor $C$, forms and integrator, and three additional analog switches ($SW_P$, $SW_S$ and $SW_{RST}$). The Skipper-CCD video signal is applied at $v_{in}$. Depending on the switches positions, its operation can be divided in four phases. In the first phase, the switches are in the position of figure 2 and the pedestal level is negatively integrated. In the second phase $SW_P$ is connected to ground and $SW_S$ to $v_{in}$ to integrate positively the signal level. At the end of the first two phases, the capacitor $C$ is charged with the value of the first pixel sample. At this point we introduce the new technique: instead of resetting $C$, we keep it charged and continue with the next pedestal-signal pair of the next sample. In this way, after pilling up $N_{skp}$ samples in the capacitor, $SW_P$ and $SW_S$ are connected to ground to sample the average of the pixel value with sub-electron noise. The last phase correspond to the reset of $C$ through $SW_{RST}$ to proceed with the readout of the next pixel. Figure 5 shows the shape of the output signal for the case of four skipper samples ($N_{skp} = 4$).

The circuit has very few components, this reduces its cost, radioactive contamination and, if the operational amplifier and switches are properly selected, it can operate at the liquid nitrogen temperature required by Skipper-CCD detectors. As we will show in the next section, due to the low sampling time (<20 ms), a multiplexer and an ADC can be used to sample several Skipper-CDS in a system with multiple Skipper-CCDs. After the readout of several Skipper-CCDs, a window of 1ms can be left to sample the output of each Skipper-CDS circuit with a multiplexer[18]. The number of channels is limited only by the multiplexer switching time. Nowadays, multiplexers with 100 ns of switching time can be found, and they can be connected in chain to increase the number of channels. The previous operation technique can also be applied to other types of detectors with non-destructive readout, like SiSeRO charge amplifiers [19] and DEPFET [20], in particular in the latest, where each detector pixel requires a readout channel.

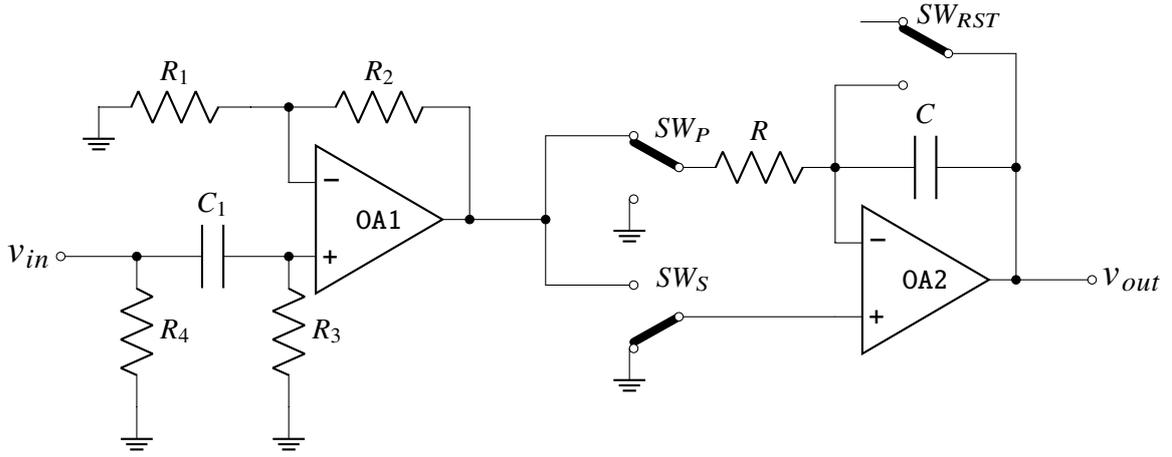

**Figure 3**. Electronic readout chain used for the experimental validation

## 3 Design and experimental verification

The Skipper-CCD used for this work, is a p-channel CCD fabricated on high resistivity, float-zone refined, n-type silicon developed by MicroSystem Labs of Lawrence Berkeley National Laboratory[21][3]. A substrate bias is applied to fully deplete the substrate, which is 675 $\mu$m thick. The high resistivity, $\approx$20 k$\Omega \cdot$ cm, allows for fully depleted operation at a substrate voltage over 70 V. The detector has 4126×866 square pixels of 15×15 $\mu$m$^2$. The Skipper-CCD vertical and horizontal registers have three-phase clocks that are designed for split readout, through the output stage at each quadrant of the sensor.

### 3.1 Design case

Between the Skipper-CCD and the Skipper-CDS we include an additional preamplifier with AC coupling. In figure 3 it is shown the full readout chain. Resistor $R_4$ is the polarization resistor of the Skipper-CCD source follower. A high pass filter for AC coupling is formed by $C_1$ and $R_3$. OA1 is in a non-inverting amplifying configuration with a gain $G_{pre}$ of x5, it improves the overall signal-to-noise ration (SNR). The OPA140 from Texas Instruments had been selected for the operational amplifiers OA1 and OA2, and the MAX333 from Maxim Integrated for the switches. Both of them had been selected mostly because they can be purchased in bare die versions to reduce radioactive contamination, and also, they had been successfully tested at liquid nitrogen temperature.

For a charge packet $Q_{pix}$, the output voltage $V_{pix}$ of the Skipper-CDS circuit is given by equation 3.1: $N_{skp}$ is the number of skipper samples, $T$ is the integration window, $G_{pre}$ is the preamplifier gain, and $RC$ is the integrator constant. The sensitivity, $S_v$, is around 1.8 $\mu V/e^-$ for the Skipper-CCD used in this work. $V_{pix}$ must be multiplied by the ADC gain to obtain the pixel value in ADU (Analog-to-digital units)

$$V_{pix} = N_{skp} G_{pre} \frac{T}{RC} S_v Q_{pix} \qquad (3.1)$$

In this work an integration window $T$ of 10 $\mu s$ has been used as design case. With the LTA it was proven that it is possible to achieve a readout noise of 0.11 $e^-$ with 400 skipper samples

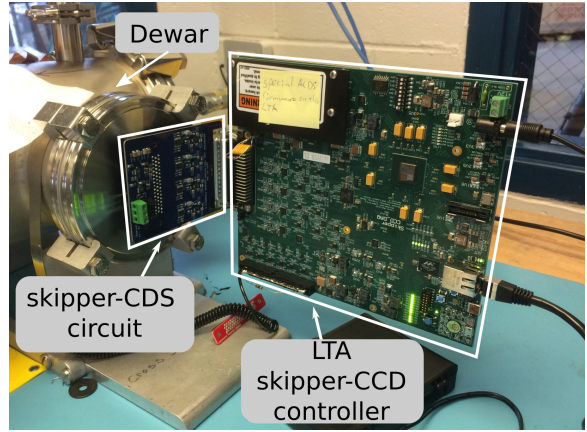

**Figure 4**. Picture of the experimental setup for testing the Skipper-CDS circuit. The Skipper-CCD is at 140 kelvin in the dewar and controlled by the LTA [12]

and 10 $\mu s$, being only limited by the noise of the source follower in the Skipper-CCD [12]. A $RC$ integration constant of 36 $\mu s$ has been selected with a 2 k$\Omega$ resistor and a 18 n$F$ capacitor. This time constant ensures the linear response of the integrator in the required 10 $\mu s$. Applying equation 3.1, for 400 skipper samples, the voltage value $V_{pix}$ for only one electron will be 1 mV. As will be shown in the following section, other integration window $T$ and number of samples $N_{skp}$, would require a proper selection of $RC$ and $G_{pre}$ to overcome the noise limitation of the operational amplifiers.

## 3.2 Experimental results

Figure 4 is a picture of the experimental setup used for testing the circuit. A dedicated PCB board was designed, it has a Skipper-CDS circuit for each quadrant of the Skipper-CCD. The LTA controller provides the necessary control signals to shift the charge in the Skipper-CCD [12]. The LTA ADCs are used to sample the output of each Skipper-CDS circuit with 18 bits in a range of 2 V, therefore, the 1 mV signal of one electron correspond to 131 ADU.

In figure 5 can be seen the output signal of the Skipper-CDS circuit digitized by the LTA ADC. In this case, only four skipper samples are piled-up in the capacitor. It is possible recognize the integration of the pedestal and signal levels of each sample, and the time period where the pixel value can be sample.

After acquiring an image with 400 skipper samples, the only processing over the image was the substraction of the mean value of each image row. In figure 6 is shown the resulting histogram of the pixel values, it is possible to recognize the single electron peaks. The peaks were fitted with Gaussian functions to extract the readout noise, giving 0.2 $e^-_{rms}/pix$

## 3.3 Noise sources

Figure 7 shows the measured noise for different numbers of skipper samples $N_{skp}$. Over 50 samples, the Skipper-CDS has a similar noise performance to the LTA system, and the noise falls with the square root of the number of samples ($\sqrt{N_{skp}}$). Below the 50 samples, the noise is dominated by the ADC noise and loss of gain. Connecting the LTA inputs to ground the ADC noise has been measured in 7 ADU, after dividing it by the system gain, it is shown in electrons units in the curve

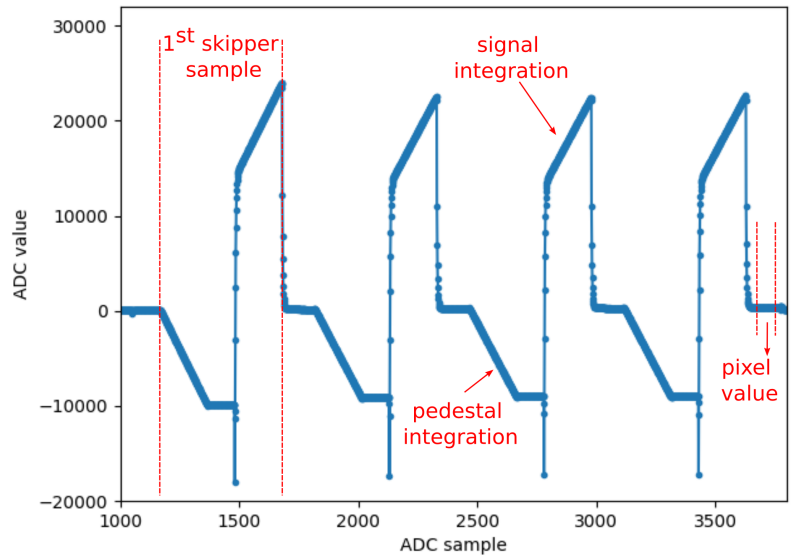

**Figure 5**. Skipper-CDS output signal for the case of four skipper samples.

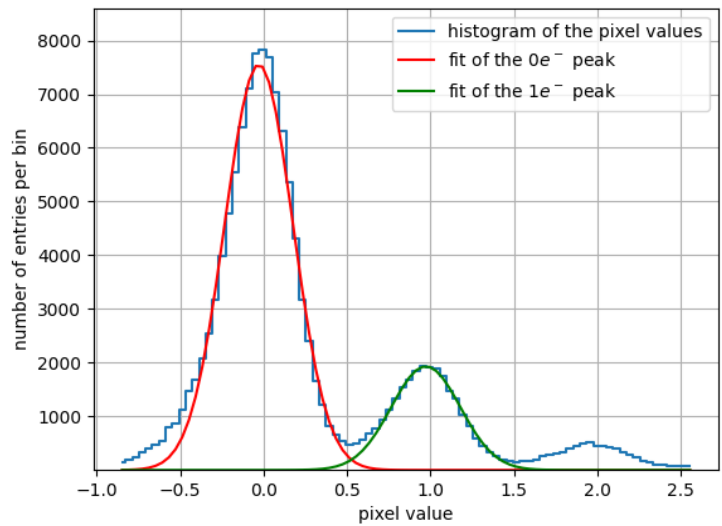

**Figure 6**. Pixel distribution after 400 samples at $10\,\mu s$ of integration time. In red is the Gaussian fit of the pixels with no charge and in green of the pixels with one electron. The resulting readout noise is $0.2\,e^{-}_{rms}/pix$

(e) of figure 7. It is important to notice that the ADC (in ADU) noise is independent of the $N_{skp}$ samples because it is added to the $V_{pix}$ output voltage of the Skipper-CDS, and therefore it is not expected to be reduced by the $\sqrt{N_{skp}}$. If only one skipper sample is acquired, the Skipper-CDS output voltage is $3\,\mu V$ that correspond to 0.32 ADU, equivalent to a total gain of $0.32\,ADU/e^-$. With 50 samples, the gain becomes $16\,ADU/e^-$, that is two times more than the ADC noise, and the noise start to be dominated by the noise from the Skipper-CCD and the operational amplifiers. To measure the contribution of the operational amplifiers, switches, and the whole readout chain, $v_{in}$ of circuit 3 was connected to ground, the result is the curve (c) of figure 7. Independently on the operation of the switches, the noise of the Skipper-CDS operational amplifier (OA2) is integrated during the entire readout time of the pixel, that involves the pile-up of the $N_{skp}$ samples. To measure this last contribution, an acquisition with $SW_p$ and $SW_s$ connected to ground was performed, the result is the curve (d) of figure 7, in where it is possible to observe that it have a small contribution to the readout noise.

The maximum number of samples that can be averaged depends on the losses of the capacitor $C$ and the reset switch $SW_{RST}$. From the MAX333 data sheet, the typical leakage current is $0.01\,nA$. As was previously mention, for the case of one sample of a charge packet of one electron, the capacitor $C$ is loaded with a voltage of $3\,\mu V$. For a leakage current of $0.01\,nA$ takes around 4.5 ms to lose the sample, this produce a fading of the samples stored in $C$ and introduce an additional noise. It is importance to notice that the readout time also includes the clocking time to shift the charge in the Skipper-CCD. In the previous experiment with 400 samples, the readout time per pixel was 17.4 ms, where 9.4 ms are consumed by the shifting of the charge in the Skipper-CCD, dominating the readout time also for a lower number of $N_{skp}$. At this point, it can be seen that the switch leakage current is the dominant noise source producing the difference from $0.2\,e^-_{rms}/pix$ to $0.11\,e^-_{rms}/pix$ respect to the LTA. Ideally, the total pixel readout time should not exceed the maximum of 4.5 ms.

From the previous analysis, it can be seen that the design of the circuit is a trade-off between several factors. We propose the following design steps. The design must start from an operation point, that includes the integration window $T$, the number of skipper samples $N_{skp}$ and a pixel readout time that are needed for a desire readout noise. An $RC$ of at least three times the integration window must be selected. The preamplifier gain must be selected as high as possible to attenuate the effect of the Skipper-CDS operational amplifier noise. The value of $C$, and its voltage value for the case of one skipper sample of one electron, is preferred to be high enough to avoid its total discharge during the pixel readout time. The maximum number of electrons that can be measure will be limited by the amplifiers power supply and the ADC span.

## 4 Conclusion

In this work an analog readout circuit for Skipper-CCD detector was presented. We introduced a new operation concept, in where the DSI capacitor is used to pile-up each skipper sample to obtain at the circuit output the pixel value with sub-electron noise. An experimental validation of the circuit was presented, achieving a noise of $0.2\,e^-_{rms}/pix$. A noise analysis of the circuit was presented. The main limitation of the circuit is leakage current from the reset switch, that can be overcome by increasing the pre-amplifier gain, increase the capacitor value and reducing the pixel readout time optimizing the shifting of the charge in the Skipper-CCD. The circuit can be applied to other types of detectors with non-destructive readout and be integrated in multi-pixel detectors. The circuit requires a very small number of components, and it significantly reduce the sampling requirements and complexity of multi Skipper-CCD cameras. In particular, it is interesting in dark matter detectors with thousands of Skipper-CCDs.

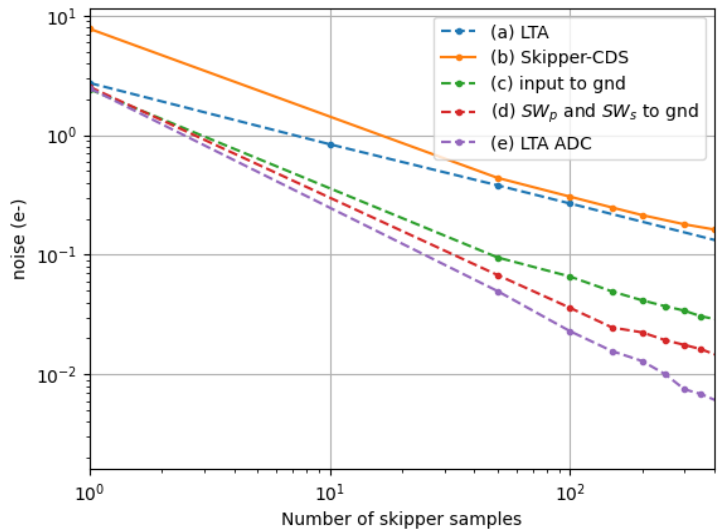

**Figure 7**. Measured noise for different numbers of skipper samples $N_{skp}$. See text for details.